\documentclass[%
pdftex
]{algotel}
\usepackage[utf8]{inputenc}
\usepackage[french]{babel}
\usepackage{graphicx}
\usepackage{caption}
\usepackage{todonotes}
\usepackage{subcaption}

\newcommand{\BATMAN}{\textsc{BATMAN}}

\author{Axel Moinet\addressmark{1}{\ }
  \and Benoît Darties\addressmark{1}{\ }
  \and Jean-Luc Baril\addressmark{1}}

\title[BATMAN : plate-forme blockchain pour l'authentification et la confiance dans les WSNs]{BATMAN : plate-forme blockchain pour l'authentification et la confiance dans les WSNs}

\address{\addressmark{1}Le2i, Université Bourgogne Franche-Comté, 9 avenue Alain Savary, 21000 DIJON, France}


\begin{document}
\maketitle

\begin{abstract}
Les réseaux sans fils (WSN) souffrent aujourd'hui d'un manque de sécurité adaptée à leurs contraintes multiples, auxquelles les solutions de gestion d'authentification et de confiance telles que PGP ne répondent que partiellement.
D'une part, les contraintes d'autonomie et de coopération des n\oe{}uds nécessaires à la garantie de la cohésion du réseau nécessitent une solution distribuée, et d'autre part les contraintes de consommation énergétique et la faible puissance de calcul des n\oe{}uds imposent l'utilisation d'algorithmes de faible complexité \cite{Zhang2014}. \`A notre connaissance, aucune solution ne permet de répondre simultanément à ces deux problématiques.
Nous proposons une nouvelle solution pour la sécurisation des WSNs nommée \BATMAN (\textit{Blockchain Authentication and Trust Module in Ad-hoc Networks}) qui répond à ces challenges. Nous présentons un modèle de gestion décentralisée pour l'authentification et la confiance, implementable sur la blockchain Tezos, et évaluons au travers de simulation les estimateurs de confiance proposés ici.
\end{abstract}

Les réseaux de capteurs sans fils (WSNs) sont au coeur du développement de l'Internet des Objets (IdO). Les solutions d'authentification et de confiance doivent s'adapter aux contraintes de ces réseaux, telles que la faible puissance de calcul des n\oe{}uds, leurs capacités matérielles ou encore la faible consommation énergétique~\cite{Medaglia2010}. Nous séparons les problématiques de sécurité sur ces réseaux en deux domaines distincts. D'une part la confidentialité et l'intégrité des données échangées, d'autre part la gestion d'authentification et de confiance des n\oe{}uds du réseau. 
Les challenges liés à l'authentification et la confiance peuvent être considérés comme les deux pendants d'un même problème. L'authentification permet d'acquérir une information sur l'identité des différents acteurs présents, tandis que la confiance nous donne une indication quant à la fiabilité et au risque de défaillance potentielle dans la communication avec un pair. Dans les réseaux traditionnels, une autorité centralisée (serveur) est en charge de ces tâches, ce qui permet d'avoir un haut niveau de sécurité et de confidentialité. Cette approche possède un inconvénient majeur, le serveur étant le point critique de sécurisation du réseau et donc de défaillance, il devient la cible privilégiée des attaques. De plus, les contraintes au sein des WSNs rendent cette approche inapplicable dans ces réseaux, notamment de part leur nature évolutive.
Une Infrastructure à Clé Publique (PKI) est habituellement utilisée comme base pour l'identification des n\oe{}uds. On peut notamment citer leur utilisation au sein de Transport Layer Security (TLS) ou encore de Pretty Good Privacy (PGP). PGP est un programme permettant aux utilisateurs de sécuriser leurs communications sans utiliser de serveur central pour la confirmation des identités~\cite{zimmermann1995official}. Pour cela, PGP utilise un réseau de confiance permettant de donner une indication de fiabilité quand à l'identité de l'interlocuteur et des actions pour lesquelles il est autorisé. Cependant, même une solution décentralisée telle que PGP n'est pas adaptée aux WSNs, car sa gestion ne peut pas être automatisée.

Il n'existe à notre connaissance aucune proposition permettant de répondre simultanément aux besoins d'authentification et de gestion de confiance et qui soit adaptée aux WSNs. Nous proposons un nouveau modèle basé sur la technologie blockchain. Ce modèle appelé \textit{\textbf{B}lockchain \textbf{A}uthentication and \textbf{T}rust \textbf{M}odule in \textbf{A}d-hoc \textbf{N}etworks} (\BATMAN) permet d'adresser les problématiques d'authentification et de gestion de confiance dans les WSNs. Ce modèle est conçu dans l'optique d'un déploiement sur la plate-forme de caméras intelligentes Wiseye\cite{wiseeye}, actuellement développée au Le2i.

\section{Blockchain Authentification and Trust Module in Ad-hoc Network}

Dans cette section, nous présentons  les principales caractéristiques de l'architecture \BATMAN~telle que nous l'avons définie.
\BATMAN~fonctionne par l'interaction entre deux grandes parties, en charge respectivement de l'authentification et de la confiance entre les n\oe{}uds.
Pour chacune de ces parties, nous présentons brièvement les différents challenges et verrous auxquels notre modèle est confronté,  ainsi que les solutions techniques apportées.
Le point critique pour l'élaboration de \BATMAN~est de disposer d'une structure de données distribuée qui soit aussi résiliente aux tentatives d'influence sur le résultat d'un calcul donné. Le meilleur exemple d'un système correspondant à ces caractéristiques sont les crypto-monnaies telles que Bitcoin\cite{nakamoto2008bitcoin}, qui utilisent la technologie blockchain pour atteindre cet objectif.
La blockchain est un registre inaltérable et distribué de données transactionnelles regroupées en blocs, ce qui permet de stocker les informations nécessaires sans nécessiter de n\oe{}ud central. Par son protocole de consensus et de transaction, elle est résiliente à l'influence et à l'intérêt particulier des n\oe{}uds qui la supportent. Enfin, la majorité des plate-formes blockchain fournissent le support de l'exécution de code distribuée pour l'automatisation des tâches au sein du réseau, à travers les smarts contracts.
Ces trois propriétés nous permettent de garantir un modèle d'évaluation neutre, déterministe et disponible à la demande, tout en restant évolutif.

\subsection{Gestion de l'authentification et de l'identité dans~\BATMAN}

Nous décrivons dans cette partie les mécanismes d'authentification et de gestion d'identité que nous avons implémenté dans \BATMAN~: nous identifions les différents n\oe{}uds du réseaux en leur associant une structure identité composée d'un triplet ($hash_{M}, hash_{UUID}, hostname)$, dans lequel $hash_{M}$ et  $hash_{UUID}$  désignent respectivement le hash de la clé publique maîtresse et d'un identifiant unique, et $hostname$ un nom d'hôte à utiliser sur le réseau.
À cette structure sont associés les hash de trois clés secondaires utilisées pour l'authentification $hash_{SA}$, la signature $hash_{SS}$ et le chiffrement $hash_{SC}$, sur un modèle équivalent à PGP. La génération des clés et de l'identifiant unique est hors du champ de ce papier.
De manière similaire à PGP, nous imposons une durée de validité maximum aux clés enregistrées pour mitiger la possibilité d'attaque par découverte de clé, mais également pour éliminer les n\oe{}uds inactifs ou ayant été retirés du réseau.

L'enregistrement et la gestion des identités est réalisée par deux smart contracts. Le premier permet de créer l'enregistrement de nouvelles identités. À l'enregistrement, un nouveau contrat dédié à la gestion des clés cryptographiques de la nouvelle identité est émis. Ce contrat permet de réaliser la mise à jour et la révocation des clés en cas de compromission ou de perte d'une clé secondaire. En cas de compromission  ou de perte de la clé maîtresse, la révocation oblige la régénération d'une nouvelle identité pour le n\oe{}ud. Le second contrat permet aux n\oe{}uds de valider leur identité au sein d'un \textit{Web of Trust} similaire à celui de PGP, dans lequel les n\oe{}uds valident les identités de leurs pairs par une signature.
Afin de mitiger la possibilité d'attaque Sybil, nous utilisons un système de \textit{Proof of Work}  qui impose une valeur numérique maximale pour $hash_{UUID}$, sur le même modèle que ce qui est utilisé au sein de la blockchain Tezos\cite{goodman2014tezos}.

\subsection{Mécanismes pour gestion de la confiance des n\oe{}uds}
\label{sub:confiance}
Notre modèle a pour ambition de répondre aux exigences de confiance en terme de fiabilité d'un n\oe{}ud réseau~\cite{Sun2006}. Notre analyse s'appuie sur la définition de la confiance de Gambetta\cite{Gambetta1990}, selon laquelle la confiance est vue comme une probabilité subjective donnée par un premier acteur quant à la réalisation correcte d'une action précise par un second acteur dans le futur. Notre approche distingue la confiance comme étant la probabilité subjective de réalisation de l'action, et la réputation comme une probabilité consensuelle, qualifiant la vision globale de l'ensemble des acteurs sur la réalisation de l'action. Nous avons inclus dans \BATMAN~les mécanismes permettant calculer les probabilités correspondant à la réputation et à la confiance d'un n\oe{}ud. Dans un soucis de concision, nous ne présentons ici que la partie du modèle liée au calcul de réputation, soit la probabilité de réalisation d'une action vue de l'ensemble du réseau.

Nous définissons une action comme une interaction entre n\oe{}uds, par exemple l'accès à une ressource. La réalisation d'une action $A$ par un n\oe{}ud $n$, notée $A_{n}$, génère un évènement noté $E_{t}^{n}$ au moment $t$ de l'action. Chaque évènement prend une valeur binaire, caractérisant l'échec ($0$) ou la réussite de l'action ($1$), et est assimilé à une variable aléatoire dans l'ensemble $[0, 1]$, suivant la loi de distribution discrète de Bernoulli.
La réalisation future du même type d'action $A_{n}$ consiste alors en un nouveau tirage. Estimer sa probabilité de réussite ou d'échec revient alors à estimer le paramètre $p$ pour la réalisation de $A_{n}$, noté $p(A_{n})$.
Nous utilisons un estimateur du maximum de vraisemblance ($ML$) pour déterminer $p(A_{n})$ à partir des évènements passés, qui, pour la loi de Bernoulli, consiste au rapport des succès sur le nombre de tirages $|E_{T}^{n}|$, où $T$ est le temps du dernier évènement. Ici, les échecs valant $0$, la somme des évènements est égale à la somme des succès, et ce calcul correspond à la moyenne des évènements sur la durée de vie du réseau $p(A_{n}) \approx ML(n, T) = \big(\sum_{t=0}^{t=T} E_{t}^{n} \big)/ {|E_{T}^{n}|}$ \label{eq:genform}. Cette définition nécessiterait cependant de disposer de tous les échantillons survenus depuis le démarrage du modèle, donc de devoir stocker un grand nombre d'échantillons, et possède une complexité algorithmique $O(t)$ non bornée donc insatisfaisante à nos besoins. Nous proposons donc trois méthodes différentes pour réaliser l'estimation et présentées dans la Table~\ref{tab:methodes}
: les deux premières méthodes utilisent pour une méthode de fenêtrage, l'une temporelle $MLT(n, T, s)$ avec $s$ taille de la fenêtre et présentée en Equation~\ref{eq:mlt}, et l'autre en nombre d'évènement $MLE(n, T, N_{e})$ avec $N_{e}$ taille de la fenêtre et présentée en Equation~\ref{eq:mle}. La dernière méthode, $MLM(n, T)$, présentée en Equation~\ref{eq:mlm}, utilise une reformulation de $ML(n, T)$ sous forme de suite. 

\vspace*{-0.3cm}
\begin{table}[ht]
\begin{tabular}{p{0.4\textwidth} p{0.55\textwidth}}
\vspace*{-0.3cm}
\begin{equation} \label{eq:mlt}
p(A_{n}) \approx MLT(n, T, s) = \frac{\sum_{t=s}^{t=T} E_{t}^{n}}{|E_{T}^{n} \setminus E_{s}^{n}|}
\end{equation} 
\vspace*{-0.3cm}& 
\vspace*{-0.3cm}
\begin{equation} \label{eq:mle}
p(A_{n}) \approx MLE(n, T, N_{e}) = \frac{\sum_{t=s}^{t=T} E_{t}^{n}}{{|E_{T}^{n} \setminus E_{s}^{n}|}}, {|E_{T}^{n} \setminus E_{s}^{n}|} = N_{e}
\end{equation}
\end{tabular}
\vspace*{-0.3cm}\begin{center}
\begin{tabular}{p{0.8\textwidth}}
\begin{equation}\label{eq:mlm}
p(A_{n}) \approx MLM(n, T) = \frac{MLM(n, T-1) * |E_{T-1}^{n}| + E_{T}^{n}}{|E_{T}^{n}|}, MLM(n, 1) = E_{1}^{n}
\end{equation}
\end{tabular}
\end{center}
\vspace*{-0.3cm}
\caption{Méthodes proposées pour réaliser l'estimation de la réputation dans \BATMAN}
\label{tab:methodes}
\end{table}
Ces trois méthodes permettent de borner le nombre d'échantillons ainsi que la complexité algorithmique, qui est respectivement pour chaque méthode en $O(s)$, $O(N_{e})$ et $O(1)$, ce qui remplit notre condition de compatibilité avec des n\oe{}uds à faible puissance de calcul.
Nous avons implémenté ces méthodes au sein de \BATMAN~en ajoutant aux smart contracts d'authentification un smart contract chargé de collecter les évènements pour chaque n\oe{}ud du réseau et de renvoyer son estimation de $p(A_{n})$ à la demande. Ce contract est émis lors de l'enregistrement, en utilisant un code préchargé sur la blockchain, ce qui nous permet de s'assurer que le code éxécuté ne peut être altéré.

\section{Validation du modèle de réputation par simulation}
Dans cette partie, nous évaluons au travers de simulations les trois méthodes proposées pour réaliser l'estimation de la réputation, présentées en Section~\ref{sub:confiance}. Les simulations ont été réalisées sous le logiciel GNU Octave.
Pour conduire nos simulations, nous introduisons une probabilité de défaillance des n\oe{}uds $p(A_{n})$, déterminée suivant une loi normale de moyenne $\mu = 0.5$ et de déviation standard $\sigma = 0.2$ (valeurs choisies arbitrairement).  Avec ces paramètres, il y a $68\%$ de chance que la probabilité pour un n\oe{}ud de réaliser une action avec succès soit comprise entre $\mu - \sigma$ et $\mu + \sigma$ (ici entre $0.3$ et $0.7$). 
Le trafic émis par chaque n\oe{}ud $n$ à chaque moment $t$ est caractérisé par un évènement $E_{t}^{n}$ dont la probabilité de réussite ou d'échec est $p(E_{t}^{n} = 1) = p(A_{n})$.

Nos simulations ont pour but de montrer l'intérêt de notre approche pour chacune de nos trois méthodes sur un réseau composé de $10$ n\oe{}uds, en étudiant l'influence du temps d'utilisation ainsi que des tailles de fenêtres utilisées. La Table~\ref{fig:params} décrit les principaux paramètres  de notre simulation, notamment le temps de simulation $T$, la taille de la fenêtre temporelle $s$ et la taille de la fenêtre d'évènement $N_{e}$. 
\begin{table}[ht]
\begin{center}
\begin{tabular}{c|c|c|c}
Paramètre & Valeur min & Valeur max & Incrément \\
\hline
$T$ & 500 & 5000 & 500 \\
$s$ & 100 & 300 & 25 \\
$N_{e}$ & 100 & 300 & 25 \\
\end{tabular}
\caption{Paramètres de simulation}\label{fig:params}
\end{center}
\end{table}
\vspace*{-0.4cm}
\begin{center}
\begin{figure}[htbp]
    \centering
    \begin{subfigure}[b]{0.45\textwidth} 
        \centering \includegraphics[width=\textwidth]{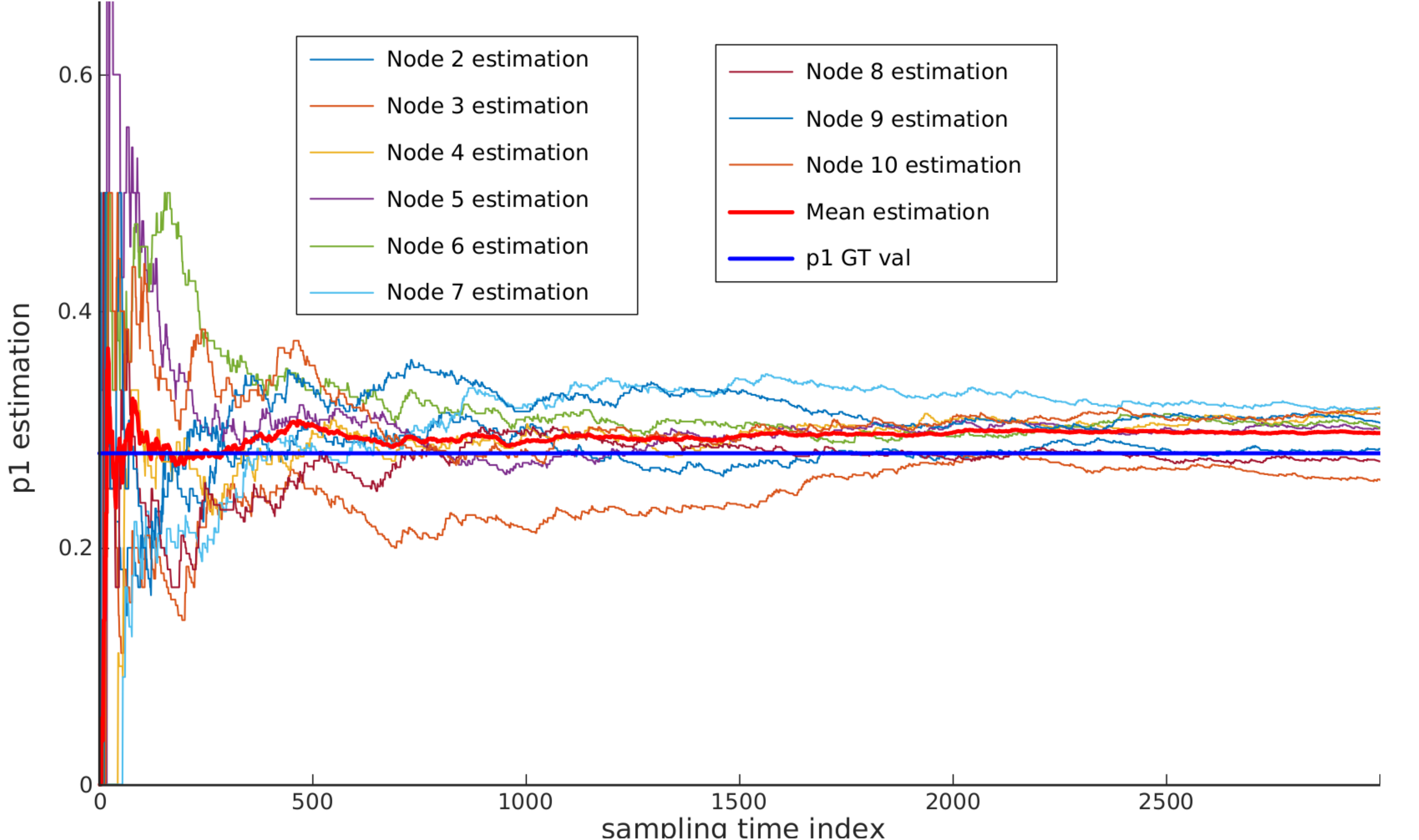}
        \caption{$p1=MLM(1, 10)$}\label{fig:MLM}
    \end{subfigure}
    ~
    \begin{subfigure}[b]{0.45\textwidth}
    \centering
    \begin{subfigure}[b]{0.5\textwidth}
        \centering \includegraphics[width=\textwidth]{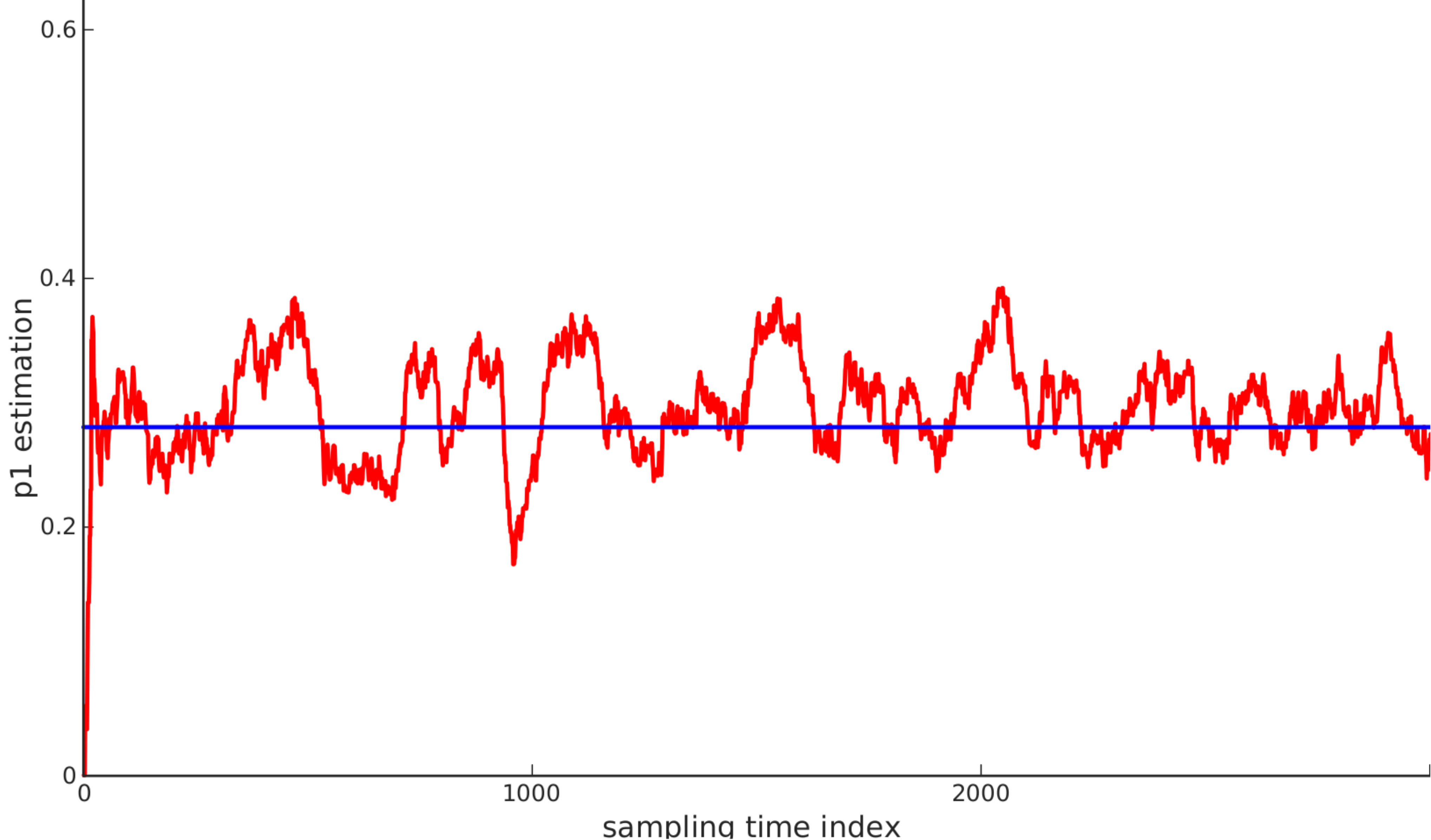}
        \caption{$p1=MLE(1, 10, 150)$}\label{fig:MLT}
    \end{subfigure}
    
    \begin{subfigure}[b]{0.5\textwidth}
        \centering \includegraphics[width=\textwidth]{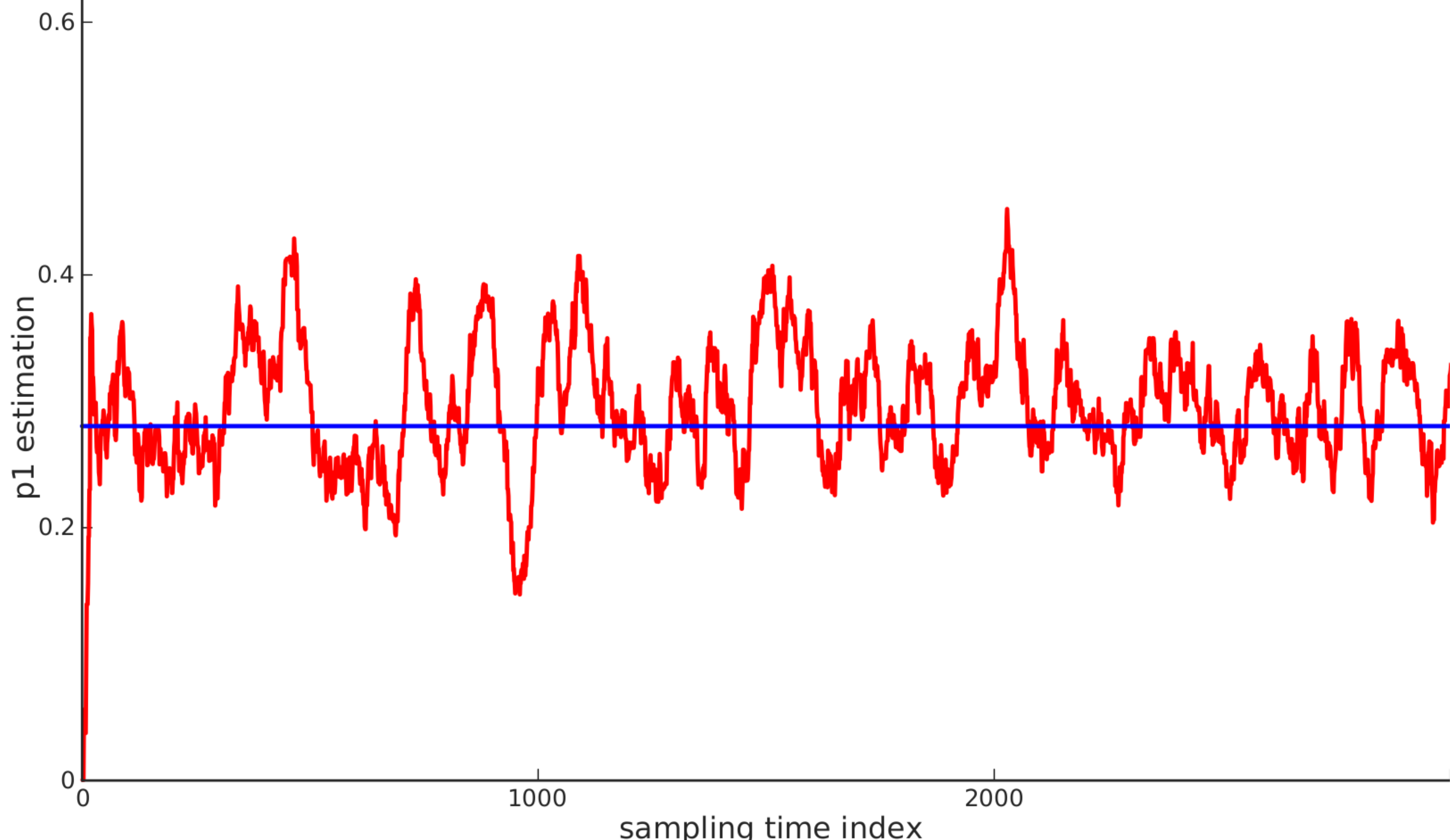}
        \caption{$p1=MLT(1, 10, 150)$}\label{fig:MLW}
    \end{subfigure}
    \end{subfigure}
    \caption{Estimations $p1$ de $p(A_{n}) = 0.28$ par $MLM$, $MLT$ et $MLW$ avec $T = 3000$}\label{fig:MLX}
\end{figure}
\end{center}
\vspace*{-0.2cm}
Nous mesurons pour le n\oe{}ud $1$ l'estimation $p1$ de $p(A_{1})$ pour les trois méthodes $MLM$, $MLE$ et $MLT$ décrites. Les résultats de ces mesures sont présentés en Figure~\ref{fig:MLX}. Ces premiers résultats montrent que la méthode $MLM$ converge vers $p(A_{n})$ avec une erreur moyenne stable, au contraire des méthodes $MLE$ et $MLT$, au comportement plus ératique, comme montré en Figure~\ref{fig:MLX} (b) et (c).

Sur l'ensemble des résultats, la méthode par fenêtre temporelle $MLT$ apparaît comme étant la moins stable dans ses estimations, ce qui était prévisible compte tenu de l'incertitude sur le nombre d'évènements $|E_{s}^{n} \setminus E_{T}^{n}|$ utilisés dans le calcul. Pour les méthodes $MLM$ et $MLE$, les résultats présentés nous permettent de définir des seuils de fenêtrage $|E_{T}^{n} \setminus E_{s}^{n}|$ pour atteindre une erreur moyenne sur l'estimation de $p(A_{n})$, de $100$ pour une erreur moyenne de $5\%$ et de $150$ pour une erreur moyenne de $3\%$.\\

Dans ce papier, nous nous sommes focalisés sur les aspects théoriques et architecturaux de notre solution. Les résultats obtenus par simulation nous permettent de passer à l'étape suivante en réalisant une implémentation de \BATMAN~sur la blockchain Tezos \cite{goodman2014tezos}, choisie pour son système de preuve \textit{Proof of Stake} qui offre des performances intéressantes en termes de consommation énergétique, et donc adaptées aux WSNs. Cette implémentation sera déployé sur un réseau de caméras intelligentes développé au Le2i pour tests en conditions réelles.
\enlargethispage{1cm}
\nocite{*}
\bibliographystyle{alpha}

\newcommand{\etalchar}[1]{$^{#1}$}

\label{sec:biblio}

\end{document}